\documentclass[prl,twocolumn,showpacs,preprintnumbers,amsmath,amssymb,floatfix]{revtex4}

\usepackage{graphicx}
\usepackage{dcolumn}
\usepackage{bm}
\usepackage{color}
\usepackage{amsmath}
\usepackage{multirow}
\usepackage{mathrsfs}

\newcommand{\KOsO}{KOs$_2$O$_6$}
\newcommand{\CsOsO}{CsOs$_2$O$_6$}
\newcommand{\RbOsO}{RbOs$_2$O$_6$}

\newif\ifetal
\etalfalse

\begin{document}

\title{\boldmath Scanning Tunneling Spectroscopy in the Superconducting State and Vortex Cores of the $\beta$-pyrochlore {\KOsO}}

\author{C. Dubois}
 \email{duboisc@mit.edu}
\author{G. Santi}
\author{I. Cuttat}
\author{C. Berthod}
\author{N. Jenkins}
\author{A. P. Petrovi\'{c}}
\author{A. A. Manuel}
\author{\O. Fischer}
\affiliation{ DPMC-MaNEP, Universit\'e de Gen\`eve, Quai Ernest-Ansermet
24, 1211 Gen\`eve 4, Switzerland}
\author{S. M. Kazakov}
\author{Z. Bukowski}
\author{J. Karpinski}
\affiliation{Laboratory for Solid State Physics ETHZ, CH-8093
Z\"{u}rich, Switzerland}

\date{\today}

\begin{abstract}
We performed the first scanning tunneling spectroscopy measurements on the
pyrochlore superconductor {\KOsO} ($T_{c}=9.6$~K) in both zero magnetic
field and the vortex state at several temperatures above 1.95~K.  This
material presents atomically flat surfaces, yielding spatially homogeneous
spectra which reveal fully-gapped superconductivity with a gap anisotropy
of 30\%. Measurements performed at fields of 2 and 6~T display a hexagonal
Abrikosov flux line lattice. From the shape of the vortex cores, we extract
a coherence length of 31--40~{\AA}, in agreement with the value derived
from the upper critical field $H_\mathrm{c2}$. We observe a reduction in
size of the vortex cores (and hence the coherence length) with increasing
field which is consistent with the unexpectedly high and unsaturated upper
critical field reported.
\end{abstract}


\maketitle

The discovery of superconductivity in the $\beta$-pyrochlore osmate
compounds AOs$_{2}$O$_{6}$ (A = K, Rb, Cs) \cite{Hiroigroup} has
highlighted the question of the origin of superconductivity in classes of
materials which possess geometrical frustration~\cite{Anderson-73,Aoki-04}.
Interest has been predominantly focused on the highest-$T_{c}$ compound
{\KOsO} which presents many striking characteristics. In particular, the
absence of inversion symmetry in its crystal structure~\cite{Schuck-06}
raises the question of its Cooper pair symmetry and the possibility of spin
singlet-triplet mixing~\cite{Frigeri-04,Hayashi-06}.

The pyrochlore osmate compound {\KOsO} displays a critical temperature $T_c
=9.6$~K, the largest in its class of materials ({\CsOsO} and {\RbOsO} which
differ only by the nature of the alkali ion have $T_c$s of 3.3 and 6.3~K
respectively). Although band structure calculations show that the K ion
does not influence the density of states (DOS) at the Fermi
level~\cite{Kunes-04,Saniz-04}, it seems to affect several key
properties~\cite{Kunes-06}.  In particular, the first order phase
transition revealed by specific heat measurements in magnetic fields at the
temperature $T_{p}\approx 7.5$~K has been ascribed to a ``freezing'' of its
rattling motion~\cite{Hiroi-06}. The negative curvature of the resistivity
as a function of temperature also indicates a large electron-phonon
scattering~\cite{Hiroi-05}. Specific heat measurements~\cite{bruhwiler-06}
suggest the coexistence of strong electron correlations and strong
electron-phonon coupling, two generally antagonistic phenomena with respect
to the superconducting pairing symmetry. The nature of the symmetry remains
a controversial subject in the literature. NMR~\cite{Arai-05} and
$\mu$SR~\cite{Koda-06} data suggest anisotropic gap functions with nodes
whereas thermal conductivity experiments~\cite{Kasahara-06} favor a
fully-gapped state.

The peculiar behavior of {\KOsO} is demonstrated by its upper critical
magnetic field $H_\mathrm{c2}$, whose temperature dependence is linear down
to sub-Kelvin temperatures and whose amplitude is above the Clogston
limit~\cite{Shibauchi-06}. One possible interpretation is the occurrence of
spin-triplet superconductivity driven by spin-orbit
coupling~\cite{Frigeri-04,Hayashi-06}. Alternatively, it has also been
suggested that this behavior can be explained by the peculiar topology of
the Fermi surface (FS) sheets of {\KOsO}, assuming that superconductivity
occurs mainly on the closed sheet~\cite{Shibauchi-06}.

The understanding of the physics of this compound would greatly benefit
from a detailed knowledge of the local density of states (LDOS). Scanning
Tunneling Spectroscopy (STS) is an ideal tool for this, particularly since
it allows one to map the vortices in real space and also access the normal
state below $T_c$ by probing their
cores~\cite{Hess-89,DeWilde-97,Eskildsen-02,Bergeal-06}. In this Letter we
present a detailed STS study of {\KOsO} single crystals, including the
first vortex imaging in this material.

The {\KOsO} single crystals were grown from Os and KO$_2$ in oxygen-filled
quartz ampoules. Their dimensions are around 0.3 $\times$ 0.3 $\times$ 0.3
mm$^{3}$. The details of their chemical properties as well as their growth
conditions can be found in Ref.~\onlinecite{Schuck-06}. AC susceptibility
measurements show a very sharp superconducting transition ($\Delta T_{c}=
0.35$~K). Our measurements are carried out using a home-built low
temperature scanning tunneling microscope featuring a compact
nanopositioning stage~\cite{Dubois-06} to target the small-sized
crystals. Electrochemically etched iridium tips are used for STS
measurements on as-grown single crystal surfaces and the differential
conductivity was measured using a standard AC lock-in technique.

\begin{figure}[tb]
\includegraphics [width=\linewidth,clip] {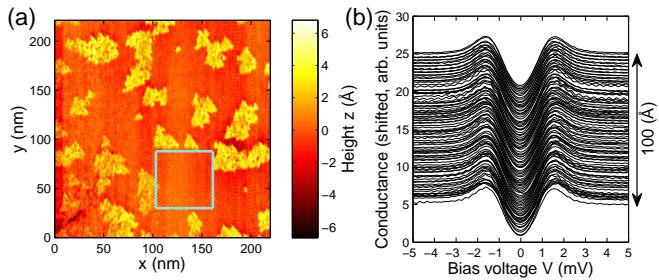}
\caption{\label{Fig_1}(a) Large-scale topography of {\KOsO} ($T=2$ K,
$R_{t}= 60$ M$\Omega$); the box shows the measurement area for the vortex
maps. (b) Spectroscopic trace along a 100~{\AA} path taken on an atomically
flat region with one spectrum every 1~{\AA}.  The spectra show raw data
offset vertically for clarity ($T=2$ K, $R_{t}= 20$ M$\Omega$). }
\end{figure}

The surface topography of as-grown samples (Fig.~\ref{Fig_1}a) reveals
atomically flat regions speckled with small corrugated islands a few
{\AA}ngstr\"{o}ms high whose spectroscopic characteristics are noisy and
not superconducting (thus restraining our field of view for spectroscopic
imaging). The large flat regions display highly homogeneous superconducting
spectra (Fig.~\ref{Fig_1}b), which were perfectly reproducible over the
timescale of our experiments (4 months). We have checked that the spectra
obtained by varying the tunnel resistance $R_{t}$ all collapse onto a
single curve, thus confirming true vacuum tunneling conditions. We have
also verified that the numerical derivative of the tunnel current with
respect to the voltage gives the same spectroscopic signature as the
$dI/dV$ lock-in signal. We stress that all measurements presented in this
paper are raw data.

\begin{figure}[tbh]
\includegraphics[width=\linewidth,clip]{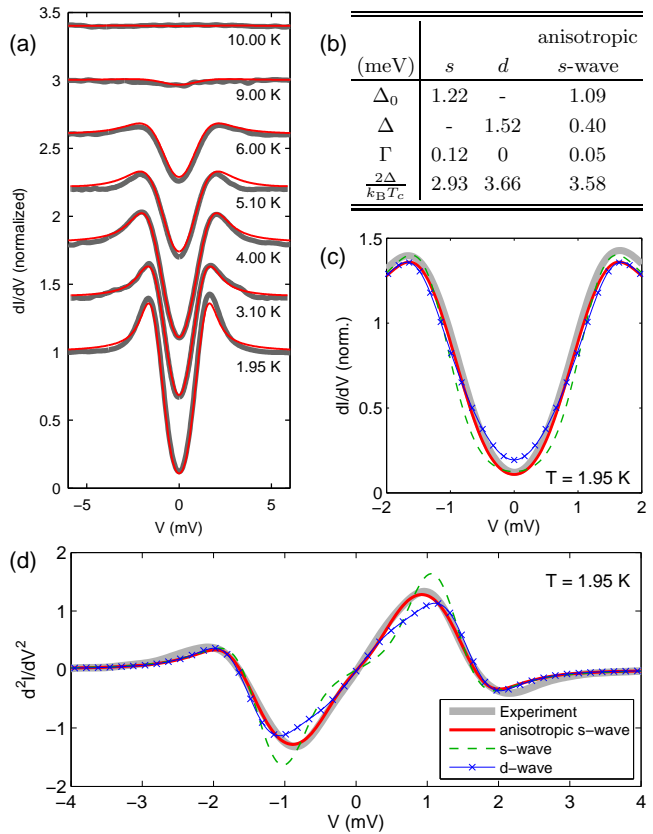}
\caption{ \label{Fig_2} 
Experimental and theoretical tunneling spectra. (a)
Normalized $dI/dV$ spectra at different temperatures from 1.95 to 10~K
(spectra are offset vertically for clarity). (b) Parameters for the
different theoretical models. (c) Comparison of the experimental spectrum
at low temperature and low energy with the different theoretical models;
the color codes are explained in (d). (d) Same as (c) for the second
derivative $d^2I/dV^2$.
}
\end{figure}

The lack of inversion symmetry in this compound together with several
experimental findings raises the question of the symmetry of the gap
function. In order to clarify this point, we have fitted our data to
several symmetry models, focusing on the question of the presence or
absence of nodes and the amplitude of any possible gap anisotropy. We
therefore considered three scenarii with an approximate angular dependence
of the gap, i.e. an isotropic $s$-wave ($\Delta_0$), a $d$-wave ($\Delta
\cos 2\phi$) with nodes and an ``anisotropic'' $s$-wave ($\Delta_0 + \Delta
\sin \phi$) which has the same angular dependence as the $s$-$p$-wave
singlet-triplet mixed state~\cite{Hayashi-06}. We do not take the real
topology of the FS~\cite{Kunes-04} into account, since it comprises two 3D
Fermi sheets and is hence unlikely to have any significant effect on the
gap structure. For an anisotropic gap, $\Delta(\phi)$, the quasiparticle
DOS is given by
$N(\omega) \propto |\mathrm{Re} [ \langle
(\omega + i \Gamma)/
\sqrt{ (\omega + i \Gamma)^2 - |\Delta(\phi)|^2 }
\rangle_\phi ]|$
where $\Gamma$ is a phenomenological scattering rate. In addition, we
included broadenings due to the experimental temperature and the lock-in in
our fits. The results are presented in Fig.~\ref{Fig_2}. The $d$-wave model
can be rejected at this stage since its zero bias conductance (ZBC) is
larger than in experiment (increasing $\Gamma$ in the model can only
increase the ZBC). The differences between symmetries appear much more
clearly in the second derivative spectrum ($d^2I/dV^2$, Fig.~\ref{Fig_2}d)
which is not surprising as it emphasizes the variations of the DOS on a
small energy scale and is very sensitive to the model parameters (in
contrast with the $dI/dV$ curve). The best fit is clearly given by the
``anisotropic'' $s$-wave model with an anisotropy of around 30\%.  With
respect to the singlet-triplet mixed state, we note that we do not see any
evidence in our data for a second coherence peak arising from spin-orbit
splitting. Since the 3D nature of both sheets implies that tunneling takes
place in both of them, the absence of a second peak also rules out the
possibility of two different isotropic gaps on separate FS sheets. Our
results would however be compatible with multiband superconductivity with
two (overlapping) anisotropic gaps. Finally, we see no signature of a
normal-normal tunneling channel in our junction, suggesting that all
electrons involved in the tunneling process come from the superconducting
condensate.

To investigate the temperature evolution of the quasiparticle DOS, we
acquired tunneling conductance spectra at different temperatures between
1.95~K and 10~K (Fig.~\ref{Fig_2}a). The closure of the gap at the bulk
$T_{c}$ shows that we are probing the bulk properties of {\KOsO}. This is
further confirmed by the fact that similar spectra were also obtained on
freshly cleaved surfaces. The totally flat conductance spectra at higher
temperature show no support for a pseudogap in the DOS above $T_{c}$,
implying that the steep decrease in the $1/(T_1T)$ curve around 16~K in NMR
data~\cite{Arai-05} must have a different origin. The spectra taken between
6 and 9~K (not shown) were very noisy. 
This could be explained by the proximity to the first order transition at
$T_p \simeq 7.5$~K~\cite{Hiroi-06}.

The BCS coupling ratio $2 \Delta_\mathrm{max}/k_\mathrm{B}T_{c}$ inferred
from our measured gaps and critical temperature is about 3.6 for the
anisotropic $s$-wave case, a value slightly smaller than that reported
from specific heat measurements~\cite{bruhwiler-06}.
However, we stress that STS is a direct probe of the superconducting gap.
Our findings therefore lead us to the conclusion that {\KOsO} is fully gapped
with a significant anisotropy of around 30\%.

\begin{figure}[tb]
\includegraphics[width=\linewidth,origin=ct]{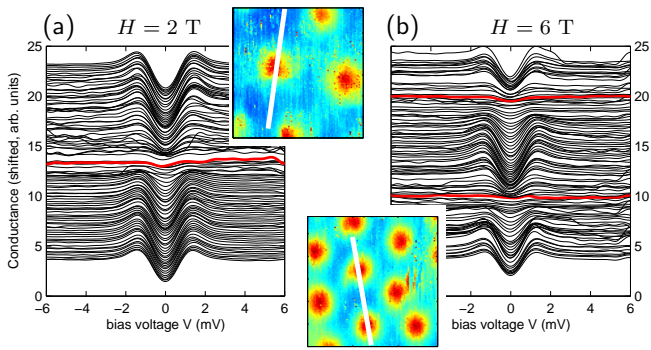}
\caption{\label{Fig_3} 
Spectroscopic traces at $T=2$~K across vortices for a field of 2~T (a) and
6~T (b). The spectra at the vortex centers are highlighted in red. The
spatial variation of the conductance is shown in the corresponding insets.
}
\end{figure}

We now focus on measurements performed in an applied magnetic field. In the
vortex cores whose radial size is roughly given by the coherence length
$\xi$, superconductivity is suppressed leading to a drastic change in the
LDOS which can be measured by STM. Our measurements were performed for two
fields, 2 and 6~T, over the particularly flat region of about $60\times
60$~nm$^2$ (Fig.~\ref{Fig_1}a). Each measurement was taken at 2~K with a
typical acquisition time of 40 hours.

\begin{figure}[tb]
\includegraphics[width=\linewidth]{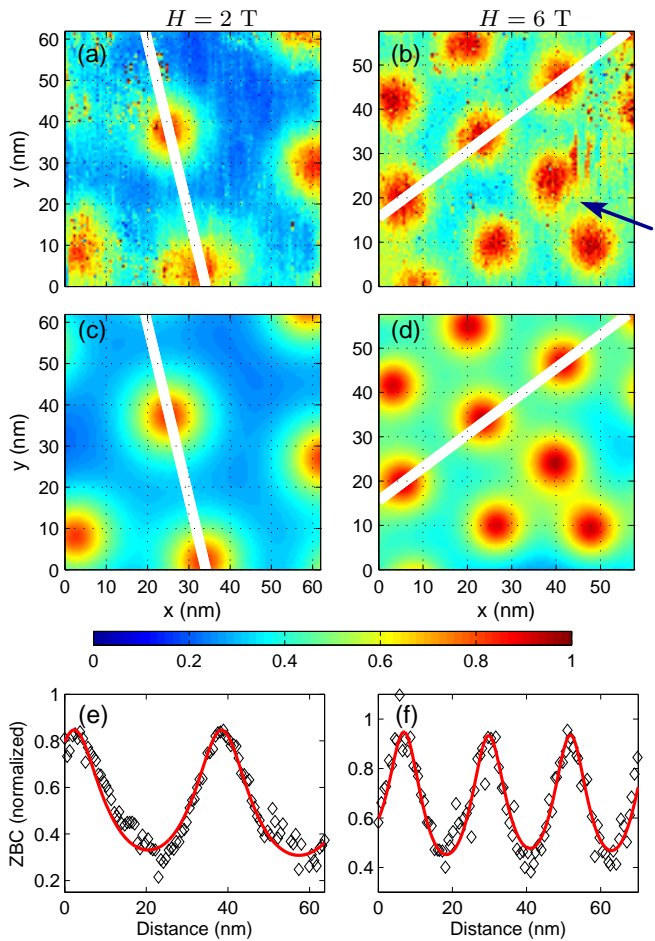}
\caption{\label{Fig_4} (a), (b) Experimental ZBC maps ($T = 2$~K)
normalized to the background conductance at 2 and 6~T respectively with
corresponding fits (c), (d); large values (red) correspond to normal
regions (i.e. vortex cores) and low values (blue) to superconducting
(gapped) regions. (e), (f) Experimental ZBC profiles across vortex centers
together with the corresponding profiles from the 2D fits (red lines).}
\end{figure}

The results are presented in Figs~\ref{Fig_3} and \ref{Fig_4}. The vortex
maps (insets of Fig.~\ref{Fig_3} and Figs~\ref{Fig_4}a and ~\ref{Fig_4}b)
show the ZBC normalized to the conductance at 6~meV. Fig.~\ref{Fig_3}
displays the spectra taken along traces passing through vortex cores for
each of the two fields considered. The suppression of superconductivity and
its effect on the conductance in a vortex core can clearly be seen. The
vortex maps show a roughly hexagonal vortex lattice with vortex spacings
$d=352 \pm 17$~{\AA} and $216 \pm 21$~{\AA} at 2 and 6~T respectively, in
agreement with the spacings $d=\left( 2\Phi_{0}/H\sqrt{3} \right)^{1/2}$
expected for an Abrikosov hexagonal lattice~\cite{Abrikosov-57},
i.e. 345~{\AA} and 199~{\AA}. We ascribe the variations in the core shapes
and the deviation from a perfectly hexagonal lattice to vortex pinning.  In
particular, the vortex identified by the arrow in Fig.~\ref{Fig_4} appears
to be split. We attribute this to the vortex oscillating between two
pinning centers during the measurement, a situation which has been seen in
other compounds~\cite{Hoogenboom-00}.  One should also note that the
islands (surface defects) at the border of the measurement area
(Fig.~\ref{Fig_1}) could influence the vortex core shapes and positions.

In order to estimate the coherence length $\xi$ from our measurements, we
now consider the spatial dependence of the ZBC. Due to the proximity of
the vortices, we model the LDOS as a superposition of isolated vortex LDOS
which can be expressed as $N(\omega,\bm{r}) = \sum_n \left| u_n(\bm{r})
\right|^2 \delta(\omega-E_n) + \left| v_n(\bm{r}) \right|^2
\delta(\omega+E_n)$, where $\psi_n(\bm{r}) = \left(u_n(\bm{r}), v_n(\bm{r})
\right)$ is the wave function of the $n$th vortex core state and $E_n$ its
energy. An approximate solution for the isolated vortex was given long
ago~\cite{Caroli-64} in which the radial dependence of each
$\psi_n(\bm{r})$ consists of a rapidly oscillating $n$-dependent Bessel
function multiplied by a $\cosh^{-1/\pi}(r/\xi)$ envelope common to all
states. We therefore construct a phenomenological model for our 2D ZBC
maps, $\sigma(\omega=0,\bm{r}) \propto N(\omega=0,\bm{r})$, by retaining
the slowly varying parts of the wave functions alone, i.e.
\begin{equation}\label{cosh}
\sigma(\omega=0,\bm{r}) = \sigma_0 + \Lambda
\sum_i \left( \cosh \frac{|\bm{r}-\bm{r}_i|}{\xi}
\right)^{-\frac{2}{\pi}}
\end{equation}
where $\sigma_0=0.13$ is the residual normalized conductance at zero bias
in the absence of field (Fig.~\ref{Fig_2}c), $\Lambda$ a scaling factor,
$\xi$ the coherence length and the sum runs over all the vortices with
positions $\bm{r}_i$ in the map. Using (\ref{cosh}), we fitted $\bm{r}_i$
and $\xi$ over the entire map for each field, thus considering all imaged
vortices to determine $\xi$.

The results from the 2D fits are presented in Fig.~\ref{Fig_4}c and d in
map format and along traces selected to pass through vortex cores in
Fig.~\ref{Fig_4}e and f. The traces help to visualize the spatial extent of
the vortices and assess the (extremely high) quality of the 2D fits.  We
first observe that the normalized ZBC between the vortices is slightly
enhanced at $H=2$~T but increases strongly at $H=6$~T with respect to the
value at zero-field (Fig.~\ref{Fig_2}c), indicating a significant core
overlap. From our data taken at $T=2$~K, we obtain $\xi = 35 \pm 3$~{\AA}
and $45 \pm 7$~{\AA} at $H=6$ and 2~T respectively (the uncertainties are
estimated from the spread of the results obtained on several maps: two for
6~T and three for 2~T). Using Ginzburg-Landau theory, we extrapolate the
corresponding $T=0$ values as $\xi = 31 \pm 3$ and $40 \pm 6$~{\AA}
respectively, consistent with the value derived from
$H_\mathrm{c2}$. Furthermore, our results indicate that the vortex size
decreases with increasing field and, although at the limit of the error
bars, we believe this trend to be genuine. In addition, this finding is
consistent with the abnormally large $H_\mathrm{c2}$: if the vortices
become smaller as the field increases, the material can accommodate more
vortices before the breakdown of superconductivity, leading to a higher
upper critical field. This correlates with the observed temperature
dependence of the upper critical field.

We find that the spectra at the vortex centers are flat for both fields
(Fig.~\ref{Fig_3}), showing the presence of localized quasiparticle states
in the vortex cores. However, our spectra show no excess spectral weight at
or close to zero bias and thus no ZBCP which is the generally expected
signature of vortex core states. The absence of a ZBCP is at first glance
striking considering the large mean free path $\ell \approx 200$~nm $\gg
\xi$ in {\KOsO}~\cite{Kasahara-06}. In fact, this absence is common to
many non-cuprate superconductors, the only known exceptions being
2H-NbSe$_2$~\cite{Hess-89,Gygi-90,Renner-91} and
YNi$_2$B$_2$C~\cite{Nishimori-04}.  Although no definitive theory currently
exists to explain such an absence, a possible explanation assumes that the
scattering rate is strongly enhanced in the vortex cores.  This
interpretation is supported by our numerical solutions of the Bogoliubov-de
Gennes equations for a single vortex with an $\bm{r}$-dependent scattering
rate $\Gamma$. Furthermore, these simulations show a radial dependence of
the LDOS which is fully consistent with (\ref{cosh}).

In conclusion, we have presented the first scanning tunneling spectroscopic
measurements on superconducting {\KOsO}. The fitted spectra demonstrate
that {\KOsO} is a fully-gapped superconductor with an anisotropy of around
30\%, possibly resulting from a $s$-$p$ singlet-triplet mixed state allowed
by the lack of inversion symmetry. We have imaged hexagonal vortex lattices
matching Abrikosov's prediction for 2 and 6~T fields.  Using Caroli-de
Gennes-Matricon theory we extract a field-dependent coherence length of
31--40~{\AA}, in good agreement with the thermodynamic estimate from
$H_\mathrm{c2}$.  The absence of a zero bias conductance peak, the apparent
field dependence of $\xi$ and the precise radial dependence of the LDOS all
call for deeper exploration.

We acknowledge T. Jarlborg, M. Decroux, I. Maggio-Aprile and P. Legendre
for valuable discussions and thank P.E. Bisson, L. Stark and M. Lancon for
technical support. This work was supported by the Swiss National Science
Foundation through the NCCR MaNEP.

\ifetal

\else


\fi

\end{document}